\documentclass[aps,prb,reprint,superscriptaddress,showpacs]{revtex4-1}
\usepackage{bm,graphicx,dcolumn,mathptmx}
\DeclareMathAlphabet{\mathcal}{OMS}{cmsy}{m}{n}
\begin{document}
\title{SU(3) Dirac electrons in the {1}/{5}-depleted square-lattice Hubbard model at {1}/{4} filling} 

\author{Yasufumi Yamashita}
\email{yamasita@ge.ce.nihon-u.ac.jp}
\affiliation{College of Engineering, Nihon University, Koriyama, Fukushima
  963-8642, Japan} 

\author{Masaki Tomura}
\affiliation{Institute for Solid State Physics, University of Tokyo, Kashiwa,
  Chiba 277-8581, Japan} 

\author{Yuki Yanagi}
\affiliation{Institute for Solid State Physics, University of Tokyo, Kashiwa,
  Chiba 277-8581, Japan} 
\affiliation{Department of Physics, Faculty of Science and Technology, Tokyo University of Science, Noda, Chiba 278-8510, Japan}

\author{Kazuo Ueda}
\affiliation{Institute for Solid State Physics, University of Tokyo, Kashiwa,
  Chiba 277-8581, Japan} 

\received{1 July 2013}
\revised[revised manuscript received ]{13 September 2013}
\published[published ]{4 November 2013}

\begin{abstract}
We investigate the magnetic and metal-insulator (M-I) phase diagram of the
${1}/{5}$-depleted square-lattice Hubbard model at ${1}/{4}$ filling by the mean-field
approximation. There exist three magnetic phases of nonmagnetic (N), 
antiferromagnetic (AF),
and ferromagnetic (F) types, each realized for the large intrasquare hopping
$t_1$, intersquare hopping $t_2$, and Coulomb interaction $U$,
respectively. Within each magnetic phase, the M-I transition of Lifshitz type
emerges and, finally, six kind of phases are identified in the $U-t_1/t_2$
plane. When $t_1=t_2$, we find that the Dirac cone and nearly flat band around
the $\Gamma$ point form the SU(3) multiplet. The SU(3) effective theory
well describes the phase transitions between NI, paramagnetic-metal (PM), 
and AF phases. The NI and AFI phases are characterized by different Berry phases 
as in polyacetylene or graphene.
\end{abstract}
\maketitle

\section{Introduction}\label{sec_1}
In recent years, physics emerging from anomalous dispersion relations, such as
a complete flat band or the Dirac cone, have attracted much attention in
condensed matter physics.\cite{kane,pesin,guo,zhao} For example when the Fermi
energy coincides with the flat band, a macroscopic degeneracy in the single
electron spectrum allows valence electrons to flip their spins freely without
losing extra kinetic energy. In a naive sense, therefore, a ferromagnetism is
achieved so as to minimize the Coulomb repulsion energy, unless other
instabilities are developed with increasing of the Coulomb interaction. Such a
mechanism, so called the flat-band ferromagnetism, is occasionally realized in
tight-binding models on geometrically frustrated lattices, generalized line
graphs,\cite{lieb,mielke,kusakabe,yanagi,miyahara} and cell-construction
networks.\cite{tasaki,mielke2} 

The Dirac cone in material science is frequently discussed for the dispersion
around the K and K' points of the honeycomb lattice in
graphene,\cite{neto,fujita} where the chiral symmetry prohibits the crossing
linear dispersion to be massive.\cite{slonczewski,hatsugai} Lately, such Dirac
electrons are demonstrated to play an important role in the two-dimensional
(2D) organic zero-gap semiconductor $\alpha$-ET$_2$I$_3$\cite{katayama,tajima}
or in the surface states of 3D topological
insulators.\cite{hasen,moore} Experimentally, the Dirac-cone electronic
dispersion is really observed in the Fe-based superconductor $\rm BaFe_2As_2$
by ARPES.\cite{richard} For the theoretical side, empirically, the Dirac-cone
system seems to be commonly found in the nearest-neighbor (NN) tight binding
models on regularly depleted lattices. The honeycomb\cite{neto} and
kagom\'{e}\cite{guo} lattices are typical examples of such a system, being
regarded as the 1/3- and 1/4-depleted triangular lattices, respectively. 
The half-filled honeycomb-lattice Hubbard model is recently studied intensively 
in connection with the M-I transition at low temperature.\cite{meng,sorella} 
The stabilities of charge-ordered states are also investigated for a spinless 
fermion model on the kagom\'{e} lattice at 1/3 filling,\cite{nishimoto} 
which is commensurate due to the lattice depletion.
As for square-lattice systems, it is known that the 1/4-depleted square lattice,
so called the Lieb model,\cite{lieb} has flat-band and Dirac-cone dispersions
at the same time.  

\begin{figure}[bp]
\begin{center}
\includegraphics[width=8.7cm]{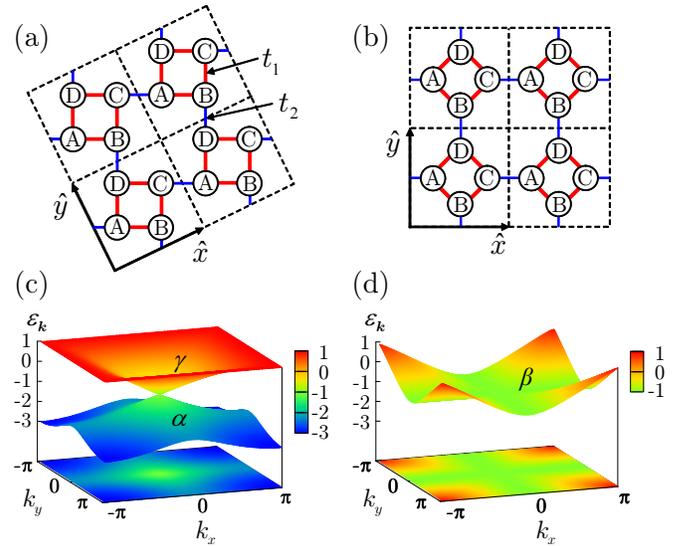}
\caption{(Color online) 
(a) The 1/5-depleted square lattice with the intrasquare ($t_1$, thick solid
  lines) and the intersquare ($t_2$, thin solid lines) hoppings. The unit
  cell, including four distinct sites A, B, C, and D, is enclosed by the
  dashed-line square connecting four vacant sites. (b) Deformed 1/5-depleted
  square lattice or decorated square lattice with C$_{\rm 4v}$ symmetry. 
  (c) Dispersion relations $\varepsilon _{\bm k}$ at $t_1=t_2=1$ for
  $\alpha$ and $\gamma$ bands, forming the Dirac cone. (d) $\varepsilon _{\bm k}$ 
  for $\beta$ band, intersecting the apex of the Dirac cone at the $\Gamma$ point. 
\label{fig_lattice}}
\end{center}
\end{figure}
In the present paper, we study the Hubbard model on the 1/5-depleted square
lattice, shown in Fig. {\ref{fig_lattice}}(a), with two different NN hoppings
of $t_1$ and $t_2$, representing the intra- and intersquare hoppings,
respectively. The original point-group symmetry of this lattice is C$_4$,
though, as long as the NN hoppings are concerned; one can deform the lattice
into the square lattice of diamonds shown in Fig. \ref{fig_lattice}(b). Thus,
the point-group symmetry is enlarged into C$_{\rm 4v}$. In the symmetric case
of $t_1=t_2=1$, diagonalizing the single-electron part of the Hamiltonian, we
find that the lowest $\alpha$ band and the third $\gamma$ band form the
characteristic Dirac cone at the $\Gamma$ point [Fig. \ref{fig_lattice}(c)]. The
apex of the cone is just located at the 1/4-filling Fermi energy and,
moreover, the second band of $\beta$ intersects this apex at the same time. In
this model, therefore, such exotic dispersions emerge simultaneously and, at
1/4 filling, the interplay of them is expected to induce intriguing physics
with the inclusion of Coulomb interaction.  

At 1/2 filling with strong electron correlations, this model is reduced to the
spin-1/2 Heisenberg model on the 1/5-depleted square lattice which describes
the spin-gap phase of $\rm CaV_4O_9$ where the plaquette spin-singlet ground
state is realized.\cite{ueda,troyer} The ratio of $J_1/J_2$ ($\propto
t_1^2/t_2^2$) controls the spin gap, and quantum phase transitions take place
between the plaquette (small $J_2$), classical AF ($J_1\sim J_2$), and dimer
(small $J_1$) phases. Also at 1/4 filling of the present study, electronic
states of plaquette or dimer structures give us an intuitive perspective to
understand the ground-state phase diagram with respect to $U$ and $t_1/t_2$,
as we will see later.  
It may be worth mentioning that Fe ions in $\rm K_{0.8}Fe_{1.6}Se_2$ are known
to form the 1/5-depleted square lattice, referred to as the
$\sqrt{5}\times\sqrt{5}$ iron-vacancy structure, with $\bm{Q}=(\pi,\pi)$
block-AF magnetic ground state.\cite{bao} 

The paper is organized as follows. In Sec. \ref{sec_2} we discuss single
electron properties focusing on the physics of the SU(3) Dirac cone. In
particular, it will become clear that the ways of symmetry lowering from the
SU(3) are crucial to give rise to M-I and magnetic transitions as a
consequence. 
After characterizing possible ordered phases in Sec. \ref{sec_2}, the
mean-field phase diagram at low temperature is presented in Sec. 
\ref{sec_3}. The NI and AFI phases are further characterized by using the
Berry phase in the latter part of this section. Finally we summarize the
conclusions and make some remarks in Sec. \ref{sec_4}. 

\section{Single electron properties}\label{sec_2}
First of all, let us investigate the noninteracting model in search of
possible metallic, insulating, and magnetic phases after including external
staggered and uniform magnetic fields. Since the Coulomb interaction is mainly
treated by the mean-field approximation in the present work, these external
fields work as the self-consistent fields reduced from the Hubbard $U$ term in
Sec. \ref{sec_3}. 
\subsection{Model and dispersion relations}
The NN tight-binding model on the 1/5-depleted square lattice is given by
${\cal H}^{(0)}=\sum_{\bm k\sigma}H^{(0)\alpha\beta}_{\bm k\sigma}c_{\bm
  k\alpha\sigma}^{\dagger}c_{\bm k\beta\sigma}$ with  
\begin{eqnarray}
\hat{H}_{\bm k\sigma}^{(0)}=
\left(
\begin{array}{cccc}
0&t_1&t_2e^{-ik_x}&t_1\cr
t_1&0&t_1&t_2e^{-ik_y}\cr
t_2e^{+ik_x}&t_1&0&t_1\cr
t_1&t_2e^{+ik_y}&t_1&0
\end{array}\right),
\label{eq_H0}
\end{eqnarray}
where $c_{\bm k\alpha\sigma}^{\dagger}$ creates a spin-$\sigma(=\pm 1)$
electron at $\alpha(=A-D)$ sublattice with the momentum $\bm k=(k_x,k_y)$. The
eigenvalue equation for Eq. (\ref{eq_H0}) is given by $F_{\bm
  k}(\varepsilon)=0$ with 
\begin{eqnarray}\hskip-1pc
F_{\bm{k}}(\varepsilon)=
\left(\varepsilon^2-t_2^2\right)^2
\!\!-\!4t_1^2
\left(\varepsilon+t_2\cos{k_x}\right)
\left(\varepsilon+t_2\cos{k_y}\right).
\label{eq_ev0}
\end{eqnarray}
The sign reversal of $t_1$ or $t_2$ is irrelevant, though the latter shifts
the momentum by $(\pi,\pi)$, therefore, we assume that $t_1$ and $t_2$ are
positive definite. For $t_1\ll t_2$ and thus $\varepsilon\simeq \pm t_2$, at
1/4 filling an electron occupies an antibonding orbital on every $t_2$ bond,
which means the PM ground state. On the other hand, for $t_1\gg t_2$ and
$\varepsilon \simeq -2t_1, 0, 0, 2t_1$, two electrons are confined within a
single $t_1$ square and the B$_1$ orbital of Eq. (\ref{eq_B1}) is doubly
occupied, resulting in the NI ground state. When $t_1=t_2$, around the $\Gamma$
point the functional forms of the Dirac cone ($\alpha$ and $\gamma$ bands) and
the $\beta$ band are, respectively, given by 
\begin{eqnarray}
\varepsilon^{(\alpha/\gamma)}_{\bm k}&\simeq& -t_1\left(1\pm \sqrt{\frac{k_x^2+k_y^2}{2}}\right),\\
\varepsilon^{(\beta)}_{\bm k}&\simeq& -t_1\left\{1-\frac{k_x^2k_y^2}{2\left(k_x^2+k_y^2\right)}\right\}, 
\end{eqnarray}
as shown in Figs. \ref{fig_lattice}(c) and \ref{fig_lattice}(d). In particular, the bottom of
the $\beta$ band along the $k_x$ and $k_y$ axes is completely flat and the system is
metallic at 1/4 filling. This is in strong contrast to the standard SU(2)
Dirac cone which is a zero-gap semiconductor in a 2D system. 

\begin{figure}[htbp]
\begin{center}
\includegraphics[width=8.6cm]{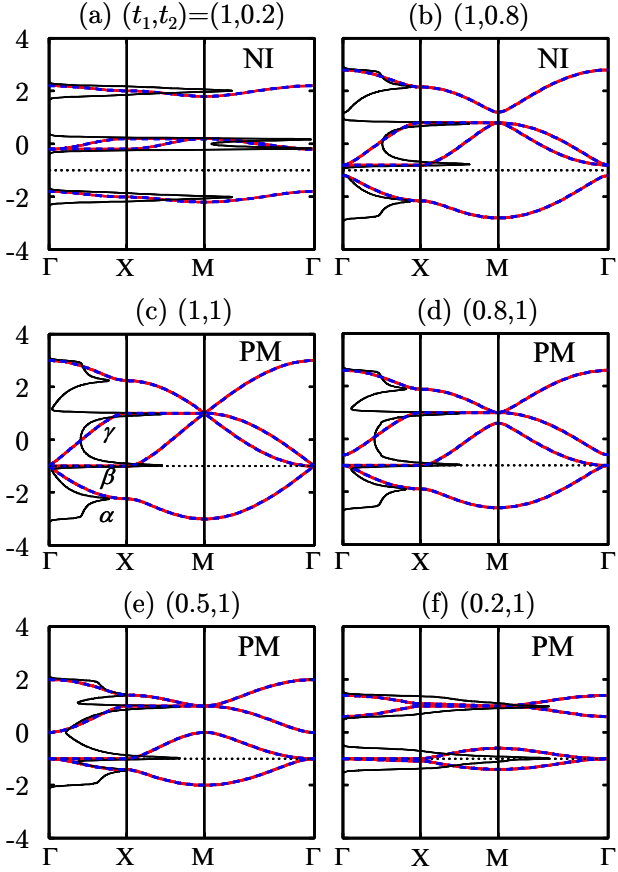}
\caption{(Color online) Dispersion relations for spin-up (dashed lines) and
  -down (solid lines) electrons along symmetry lines for various
  $(t_1,t_2)$'s. In the metallic cases of (c), (d), (e), and (f), the
  flat bands along $\Gamma$-X match the Fermi energy, depicted by the dotted
  line. The densities of states are also displayed. 
\label{fig_disp}}
\end{center}
\end{figure}

For general $(t_1,t_2)$'s, the dispersion relations along
$\Gamma$-X$(\pi,0)$-M$(\pi,\pi)$-$\Gamma$ are displayed in
Fig. {\ref{fig_disp}}. As shown in Fig. \ref{fig_disp}(c), the Dirac cone
appears also at the M point, since the relation 
\begin{eqnarray}
F_{\bm k+(\pi,\pi)}(-\varepsilon)=F_{\bm k}(\varepsilon)
\end{eqnarray}
holds due to the chiral symmetry. Along the $\Gamma$X line, putting $k_y=0$ in
Eq. (\ref{eq_ev0}), a completely flat dispersion emerges at
$\varepsilon=-t_2$. When the 1/4-filling Fermi energy hits this flat
dispersion, for $t_1\le t_2$ as shown in Figs. \ref{fig_disp}(c)-\ref{fig_disp}(f), the
ground state is PM. On the other hand, for $t_1>t_2$, a finite band gap is
open at the $\Gamma$ point as shown in Figs. \ref{fig_disp}(a) and \ref{fig_disp}(b). Lifting of
the degeneracy at the $\Gamma$ point seems to be the key to understanding the M-I
transition. 

\begin{figure}[htbp]
\begin{center}
\includegraphics[width=8.6cm]{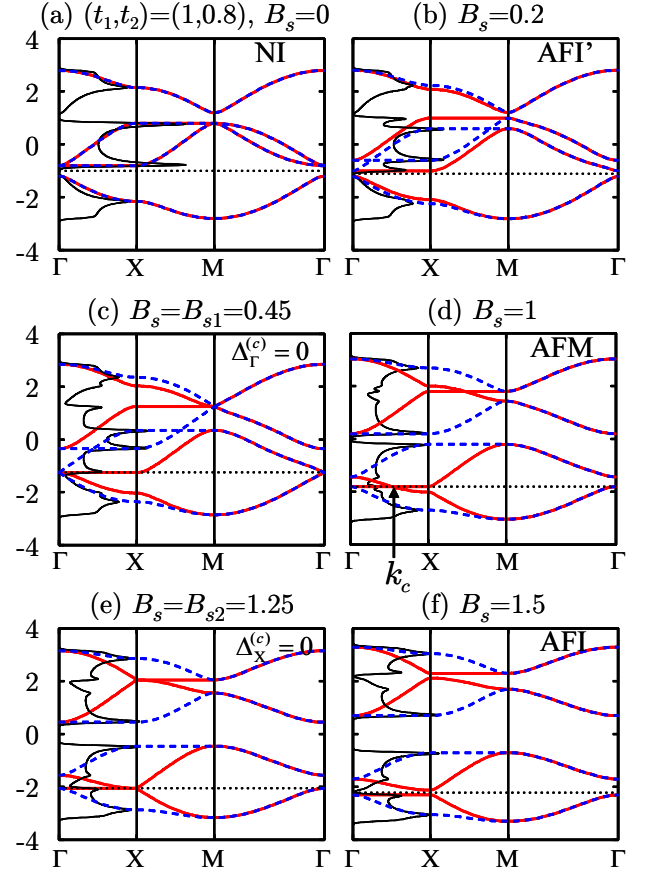}
\caption{(Color online) Dispersion relations for $(t_1,t_2)=(1,0.8)$ with
  increasing $B_s$ from 0 to 1.5, where symbols and lines are the same as in
  Fig. \ref{fig_disp}. At $\bm k=(k_c,0)$, as shown in (d), the flat band is
  intersected by another band. 
\label{fig_disp_Bs}}
\end{center}
\end{figure}

\subsection{SU(3) effective theory}
At the $\Gamma$ point, when $t_1=t_2$, Eq. (\ref{eq_H0}) is diagonalized into
diag($3t_1,-t_1,-t_1,-t_1$) by using the C$_{\rm 4v}$ basis representation of 
\begin{eqnarray}
|{\rm A_1}\rangle&=&\frac{1}{2}(c_{A}^{\dagger}+c_{B}^{\dagger}+c_{C}^{\dagger}+c_{D}^{\dagger})|0\rangle,\\
{\rm \left(|E_x\rangle,|E_y\rangle\right)}&=&\frac{1}{\sqrt{2}}\left(c_{A}^{\dagger}-c_{C}^{\dagger}, c_{B}^{\dagger}-c_{D}^{\dagger}\right)|0\rangle,\\
{\rm |B_1\rangle}&=&\frac{1}{2}(c_{A}^{\dagger}-c_{B}^{\dagger}+c_{C}^{\dagger}-c_{D}^{\dagger})|0\rangle,\label{eq_B1}
\end{eqnarray} 
where $\bm k$ and $\sigma$ indices are abbreviated.
The $\rm |A_1\rangle$ state is energetically separated from the degenerate
SU(3) multiplet. Therefore, expanding Eq. (\ref{eq_H0}) for small $\bm k$ and
$\mu=t_2/t_1-1$, transforming it into the C$_{\rm 4v}$ basis, and tracing out
the $\rm |A_1\rangle$ state, the effective SU(3) Hamiltonian is constructed
for fixed $\bm k$ and $\sigma$ as 
\begin{eqnarray}
{\cal H}^{(0)}_{\rm eff}=
-\left(1+\frac{\mu}{3}\right)I_3-\frac{2\mu}{\sqrt{3}}\lambda_8+\frac{k_x}{\sqrt{2}}\lambda_5-\frac{k_y}{\sqrt{2}}\lambda_7
\label{eq_Heff}
\end{eqnarray}
in the unit of $t_1$. Here, $I_3$ and $\lambda$'s are, respectively, the
identity and the Gell-Mann matrices of the SU(3) subspace spanned by $\rm
(|E_x\rangle, |E_y\rangle, |B_1\rangle)$. 

In the absence of $\mu$, ${\cal H}^{(0)}_{\rm eff}$ can be viewed as the
massless Dirac equation embedded in the SU(3) space. 
From the C$_{\rm 4v}$ point group symmetry, E and B$_1$ states are always 
degenerate to form the Dirac point. 
Including the mass term
proportional to $\lambda_8=(\rm|E_x\rangle\langle E_x|+|E_y\rangle\langle
E_y|-2|B_1\rangle\langle B_1|)/\sqrt{3}$, the accidental degeneracy between E-
and B$_1$-symmetric states at the $\Gamma$ point is lifted, and a positive
(negative) $\mu$ increases the energy of B$_1$(E) state. As a result, Lifshitz
transition\cite{lifshitz} separates the nonmagnetic phase into NI ($t_1>t_2$)
and PM ($t_1\le t_2$) phases, which is consistent with the simple M-I
transition picture between the isolated-square ($t_2=0$, insulating) and
isolated-bond ($t_1=0$, metallic) limits. 

\subsection{Phase diagram under a staggered magnetic field}
In order to lift the degeneracy within the E states at the $\Gamma$ point, we need
a $\lambda_3=\rm|E_x\rangle\langle E_x|-|E_y\rangle\langle E_y|$ term besides
${\cal H}^{(0)}_{\rm eff}$, which can open the charge gap for $t_2/t_1\ge1$ or
close it for $t_2/t_1<1$. For that purpose, we apply the $\bm Q= \bm 0$
staggered magnetic field defined by 
\begin{eqnarray}
{\cal H}^{(B_s)}=\sum_{{\bm k}\sigma}\sigma B_s\left(n_{kA\sigma}-n_{kB\sigma}+n_{kC\sigma}-n_{kD\sigma}\right),
\end{eqnarray}
where $n_{k\alpha\sigma}=c^{\dagger}_{k\alpha\sigma}c_{k\alpha\sigma}$. For
small $B_s$, tracing out the $|\rm A_1\rangle$ state in the same way, we can
easily show that the effective SU(3) model for ${\cal H}^{(B_s)}$, with fixed
$\bm k$ and $\sigma$, is given by
\begin{eqnarray}
{\cal H}^{(B_s)}_{\rm eff}=
\sigma B_s\lambda_3^{(\sigma)},
\end{eqnarray}
where $\lambda_3^{(\sigma)}$ is the similarly defined Gell-Mann matrix in the
spin-$\sigma$ sector. Therefore the $B_s$ term breaks the E symmetry at
the $\Gamma$ point, which induces M-I phase transitions accompanied by the $\bm
Q=\bm 0$ AF spin order. 

The eigenvalue equation for ${\cal H}^{(0)}+{\cal H}^{(B_s)}$ is now written
as $F^{(+)}_{\bm k}(\varepsilon)\cdot F^{(-)}_{\bm k}(\varepsilon)=0$ with 
\begin{eqnarray}\hskip-1pc
F^{(\sigma)}_{\bm{k}}(\varepsilon)=
\left\{\left(\varepsilon-\sigma B_s\right)^2-t_2^2\right\}
\left\{\left(\varepsilon+\sigma B_s\right)^2-t_2^2\right\}\nonumber\\
-4t_1^2
\left(\varepsilon-\sigma B_s+t_2\cos{k_x}\right)
\left(\varepsilon+\sigma B_s+t_2\cos{k_y}\right).
\label{eq_ev}
\end{eqnarray}
Solving Eq. (\ref{eq_ev}), we find the M-I phase diagram\cite{comment4} 
in the $B_s-t_1$ or $B_s-t_2$ plane as shown in Fig. \ref{fig_MI_Bs}. 
For $B_s\ne 0$, there are two kinds of insulating phases: AFI' (for small
$t_2$) and AFI (small $t_1$). In the AFI' phase, like
Fig. {\ref{fig_disp_Bs}}(b), the charge gap $\Delta^{(c)}_{\bm k}$ opens at
$\Gamma$, which is estimated from Eq. (\ref{eq_ev}) to be 
\begin{eqnarray}
\Delta^{(c)}_{\Gamma}=-2t_2-B_s+\sqrt{4t_1^2+B_s^2}.
\end{eqnarray}
As shown in Fig. {\ref{fig_disp_Bs}(c), the charge gap closes at 
$\Delta^{(c)}_{\Gamma}=0$, which determines the I-M transition line; 
\begin{eqnarray}
B_{s1}(t_1,t_2)={\left(\frac{t_1^2}{t_2}-t_2\right)}\theta(t_1-t_2).\label{eq_Bs1}
\end{eqnarray}
Once entering the metallic phase, like Fig. {\ref{fig_disp_Bs}}(d), the
complete flat band along $\Gamma$-X is intersected by the dispersive band at
$k_x=k_c$. The flat band is a localized antibonding orbital of B and D sites
with $\varepsilon=-t_2-B_s$, which must be the double root of
Eq. (\ref{eq_ev}) at $\bm k=(k_c,0)$ to give, 
\begin{eqnarray}
\hskip-1pc k_c(t_1,t_2,B_s)=2\arcsin{\sqrt{\frac{B_s}{t_2}\left\{\frac{t_2^2}{t_1^2}\left(1+\frac{B_s}{t_2}\right)-1\right\}}}.\label{eq_kc0}
\end{eqnarray}
In Fig. \ref{fig_disp_Bs}(d) of $(t_1,t_2,B_s)=(1,0.8,1)$, for instance,
$k_c$ is calculated to be 
\begin{eqnarray}
k_c(1,0.8,1)=2\arcsin{\sqrt{0.55}}\simeq 0.53\pi. \label{eq_kc}
\end{eqnarray}
Note that the dispersive band has even parity with respect to exchanging B and
D sites. That's why there is no band splitting at $\bm k=(k_c,0)$. 

With increasing $B_s$ above $B_{s1}$, $k_c$ travels from 0
[Fig. {\ref{fig_disp_Bs}}(c)] to $\pi$ [Fig. {\ref{fig_disp_Bs}}(e)], and
finally the charge gap of the AFI phase opens at the X point
[Fig. \ref{fig_disp_Bs}(f)] with 
\begin{eqnarray}
\Delta^{(c)}_{\rm X}=t_2+B_s-\sqrt{4t_1^2+(t_2-B_s)^2}.
\end{eqnarray}
Again, $\Delta^{(c)}_{\rm X}=0$ defines the M-I transition line of
\begin{eqnarray}
B_{s2}(t_1,t_2)=\frac{t_1^2}{t_2},\label{eq_Bs2}
\end{eqnarray}
as depicted in Fig. \ref{fig_MI_Bs}.
Namely, in the shaded AFM region in Fig. \ref{fig_MI_Bs}, $k_c$ ranges from 0
to $\pi$ consistent with the relations, $k_c(t_1,t_2,B_{s1})=0$ and
$k_c(t_1,t_2,B_{s2})=\pi$, as shown by using Eqs. (\ref{eq_Bs1}),
(\ref{eq_kc0}), and (\ref{eq_Bs2}). 
$B_{s1}$ and $B_{s2}$ transition lines are of Lifshitz type, which are similar
to the PM-NI phase transition at $B_s=0$. Although  
these M-I transitions are peculiar in that Fermi surfaces of line shape along
the $k_x$ and $k_y$ axes vanish at the transition.\cite{lifshitz} 

\begin{figure}[tbp]
\begin{center}
\includegraphics[width=7.6cm]{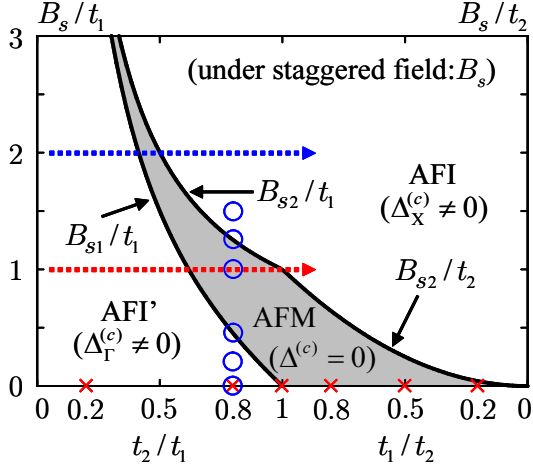}
\caption{(Color online) M-I phase diagram\cite{comment4} under the external staggered
  magnetic field ${\cal H}^{(B_s)}$. 
  All transition lines are of Lifshitz
  type, where only Fermi surface geometry changes and no other symmetry is
  broken. At the points marked by the crosses ($\times$) and open circles
  ($\circ$), respectively, the dispersion relations are shown in
  Figs. \ref{fig_disp} and \ref{fig_disp_Bs}. Along the horizontal dashed
  arrows at $B_s=t_1$ and $2t_1$, the Berry phase is calculated in
  Fig. \ref{fig_gamma}. 
\label{fig_MI_Bs}}
\end{center}
\end{figure}

\subsection{Phase diagram under a uniform magnetic field}
Figure {\ref{fig_MI_Bu}} shows the similarly derived M-I phase diagram\cite{comment4} 
under the external uniform magnetic field; 
\begin{eqnarray}
\hskip-.5pc{\cal H}^{(B_u)}=\sum_{{\bm k}\sigma}\sigma B_u\left(n_{kA\sigma}+n_{kB\sigma}+n_{kC\sigma}+n_{kD\sigma}\right).
\end{eqnarray}
In the NI phase for $t_2/t_1<1$, the charge gap is given by
$\Delta^{(c)}_{\Gamma}=2(t_1-t_2-B_u)$. With increasing $B_u$ from zero,
therefore, the NI-FM phase transition takes place at $B_{u1}=t_1-t_2$. On the
other hand, in the FM phase for $t_1/t_2\le 0.5$, the phase transition to the
fully saturated FI phase occurs at $B_{u2}=2t_1$, where $4t_1$ is the
bandwidth of the majority-spin band.

When $B_u$ is large enough compared with $t_1$ and $t_2$, the present model at
1/4 filling is reduced to the spinless 1/2-filled problem. In that case, Fermi
energy is zero because of the bipartite symmetry, and the Fermi surface, if
it exists, is defined by the equation $F_{\bm k}(\epsilon=0)=0$ resulting in
\begin{eqnarray}
\cos{k_x}\cos{k_y}=\frac{t_2^2}{4t_1^2}.
\end{eqnarray}
Therefore, $t_1/t_2=0.5$ determines the FM-FI phase transition line, as shown
in Fig. {\ref{fig_MI_Bu}} for large $B_u$. At this Lifshitz transition,
electron and hole pockets at $\Gamma$ and M points, respectively, vanish
simultaneously, see Figs. \ref{fig_disp}(d)-\ref{fig_disp}(f) at 1/2 filling of
$\varepsilon=0$.
\begin{figure}[tbp]
\begin{center}
\includegraphics[width=7.6cm]{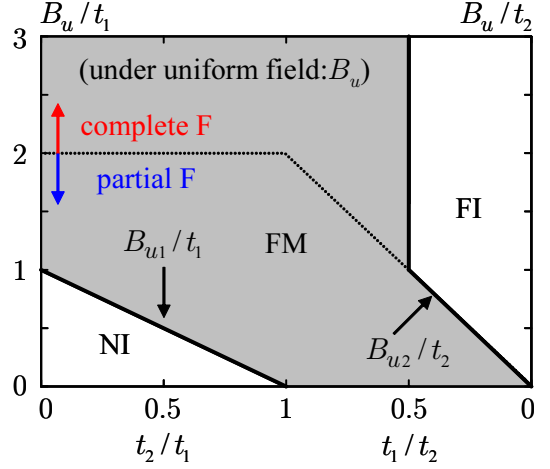}
\caption{(Color online) M-I phase diagram\cite{comment4} under the external uniform magnetic
  field ${\cal H}^{(B_u)}$. All transition lines are of Lifshitz type. The
  dotted line in the FM phase indicates the lower boundary of the
  complete ferromagnetic region. The ferromagnetic moment in FI is fully
  saturated.
\label{fig_MI_Bu}}
\end{center}
\end{figure}

\begin{figure}[bp]
\begin{center}
\includegraphics[width=7.0cm,angle=0]{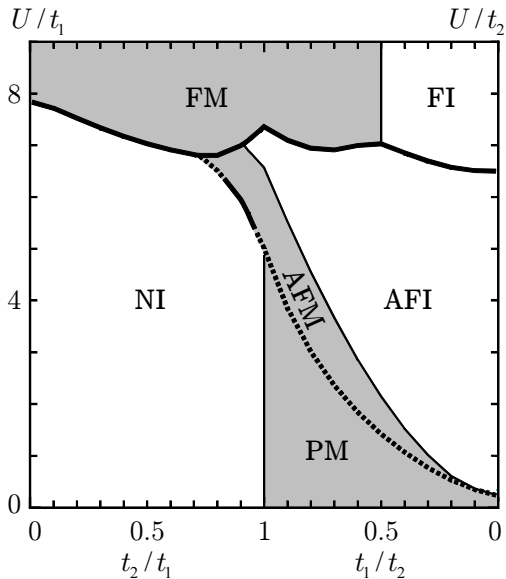}
\caption{Mean-field phase diagram in the $U-t_2$ or $U-t_1$ plane, where
  thick-solid and dashed lines represent discontinuous and continuous magnetic
  phase boundaries, respectively. The symbol AF stands for the $\bm Q=\bm 0$
  antiferromagnetic order. The ferromagnetic moment in the FM and FI is fully
  saturated. The M-I phase transition lines shown by thin solid lines are of
  Lifshitz type. 
\label{fig_pd}}
\end{center}
\end{figure}

\section{Effects of the Coulomb interaction}\label{sec_3}
\subsection{Mean-field phase diagram}
Next we consider the effect of the on-site Coulomb interaction within the
Hartree-Fock approximation, where the external fields of $B_s$ and $B_u$
discussed so far are replaced by the self-consistent fields. For numerical
stabilities, all calculations have been done at a finite temperature of
$T=0.03\times\textrm{max}(t_1,t_2)$ for the typical system size of
$200\times200$. By comparing the free energies of the nonmagnetic, 
ferromagnetic, and $\bm
Q={\bf 0}$ antiferromagnetic states, we obtain the ground-state phase diagram
as shown in Fig. {\ref{fig_pd}}. Note that temperature and the system-size
dependence of the phase boundaries are already very weak for the present
calculations. 

We have also checked the paramagnetic susceptibility $\chi_0(\bm q)$ for
$t_1<t_2$, which shows that the leading instability within the random phase
approximation (RPA) occurs at $\bm q={\bf 0}$ and the staggard moments develop
within the unit cell consistent with the $\bm Q={\bf 0}$ AF order assumed
in Fig. \ref{fig_MI_Bs}. From the largest eigenvalue of $\chi_0$, the RPA
instability is determined to be consistent with the PM-AFM transition line
shown in Fig. \ref{fig_pd}. The other simple order parameter, like $\bm
Q=(\pi,\pi)$ AF order, which is similar to the standard classical AF order in
the square lattice, is always high in energy. The other $\bm Q=(\pi,\pi)$ mode
with the ferromagnetic unit cell, that is the block-AF structure observed in
$\rm K_{0.8}Fe_{1.6}Se_2$,\cite{bao} has higher energy in this model. 

For $t_1\simeq t_2$ and small $U$, the self-consistent field does not develop
and the NI-PM Lifshitz transition at $t_1=t_2$ is straight in this region. For
$U\gg t_1$ and $t_2$, fully saturated FM and FI compete with each other, 
separated by the Lifshitz transition at $t_1/t_2=0.5$ consistent with
Fig. \ref{fig_MI_Bu}. 

On the right half of the phase diagram ($t_1/t_2\le 1$) with modest $U$, the
antiferromagnetic phases prevail over the para- and ferromagnetic
phases. Starting from the small $t_1$ limit, an electron resides on the
antibonding orbital of the $t_2$ bond, labeled by $i$, where only the spin degree
of freedom $\bm S_i$ is left.\cite{comment1} The second-order perturbation by
super-exchange processes via $t_1$ hoppings derives the effective pairwise
spin Hamiltonian, 
\begin{eqnarray}
{\cal H}_{ij}^{(2)} =\frac{t_1^2}{2}\left(\frac{1}{2t_2+U}+\frac{4}{U}\right)\bm S_i\cdot\bm S_j+\textrm{const.}
\label{eq_h2}
\end{eqnarray}
Thus, the ground state for the small-$t_1$ region should be the AF bond
spin-density wave (SDW) which is consistent with the ${\bm Q=\bm 0}$ AFI of
Fig. \ref{fig_pd}. This is because, in both ordered states, the staggard
spin order in the unit cell stretches over the entire lattice uniformly. Once
self-consistent fields are developed, there is a one-to-one correspondence
between the phase diagrams of Figs. \ref{fig_MI_Bs}, \ref{fig_MI_Bu}, and
\ref{fig_pd} by reading $B_s$ or $B_u$ as $U$ multiplied by spin
densities. Accordingly, the M-I phase boundary in the AF region in
Fig. \ref{fig_pd} is nothing but the Lifshitz transition discussed in
Fig. \ref{fig_MI_Bs}. In addition we have observed a faint trace of the AFI' phase
in a narrow region just below the NI-AFM boundary around $t_2/t_1=0.7$-1.0
(not shown in Fig. \ref{fig_pd}). 

At $t_2=0$, the model is reduced to the 4-site $t_1$-ring+$U$ problem, in
which exact FM and mean-field NI energies of $E_{\rm FM}$ and $E_{\rm NI}^{\rm
  (MF)}$, respectively, are given by 
\begin{eqnarray}
E_{\rm FM}&=&-2t_1,\\
E_{\rm NI}^{\rm (MF)}&=&-4t_1+\frac{U}{4}.
\end{eqnarray}
Therefore, $U_c=8t_1$ is found at $t_2=0$ consistent with the left end of
Fig. \ref{fig_pd}, where the finite $T$, $0.03t_1$ in the present calculation,
decreases the FM free energy and $U_c$ is slightly reduced. On the other end
of Fig. \ref{fig_pd} where $t_1$ is much smaller than $t_2$ and $U$, expanding
the FI and AFI mean-field energies as a function of $t_1$ up to the second
order, we obtain 
\begin{eqnarray}
E_{\rm FI}^{(2)}&=&-2t_2-\frac{t_1^2}{t_2},\\
E_{\rm {AFI}}^{(2)}&=&-2t_2-4t_1^2\left(\frac{1}{U}+\frac{1}{U+4t_2}\right).
\end{eqnarray}
Comparing $E_{\rm F}^{(2)}$ and $E_{\rm AFI}^{(2)}$, the mean-field $U_c$
converges to $2(1+\sqrt{5})t_2\simeq 6.47t_2$ as $t_1$ approaches zero. The
mean-field staggered moment in the AFI is estimated for small $t_1$ as 
\begin{eqnarray}
M_{\rm AFI}^{(2)}=1-8t_1^2\left\{\frac{1}{U^2}+\frac{1}{\left(U+4t_2\right)^2}\right\}.
\end{eqnarray}
For small $t_1/t_2$, in reality, the AFI region extends over $U=\infty$
because of the super-exchange interaction [Eq. (\ref{eq_h2})], as mentioned above.

The mean-field analysis 
is qualitatively justified for the Coulomb interaction $U$ smaller than the noninteracting 
bandwidth of $W=4t_1+2t_2$. When $U$ exceeds the maximum bandwidth of around 
$6t_1$ or $6t_2$, it seems that the first order transitions to fully saturated 
ferromagnetic states take place mostly irrespective of $t_1/t_2$.
This may be
an artifact due to neglecting electron correlations, though electron kinetic
energy is favorable for the spin to be aligned in parallel at 1/4 filling,
generally speaking. 

\subsection{Calculations of the Berry phase}
The AFI' (including the NI as a special case of $B_s=0$) and AFI phases are
also distinguished by using the bulk properties of Berry
phase\cite{berry,zak,resta} as a function of $k_x$, which is defined for
$L_x\times L_y$ lattice under periodic boundary conditions as follows:
\begin{eqnarray}
\gamma^{(\alpha)}_{k_x\sigma}
&=&-i\int_{-\pi}^{\pi}dk_y\langle \Phi^{(\alpha)}_{\bm k\sigma}|\frac{\partial}{\partial k_y}|\Phi^{(\alpha)}_{\bm k\sigma}\rangle\\
&=&\sum_{k_y}{\rm Im}\langle \Phi^{(\alpha)}_{\bm k\sigma}|\Phi^{(\alpha)}_{\bm k+(0,\frac{2\pi}{L_y}),\sigma}\rangle,\label{eq_gamma}
\end{eqnarray}
where $|\Phi^{(\alpha)}_{\bm k\sigma}\rangle$ is the eigenvector of the lowest
$\alpha$ band with spin $\sigma$. In the numerical calculations, we take the
unit cell as shown in Fig. \ref{fig_lattice}(b) and $L_x=L_y=512$. 

Figure \ref{fig_gamma} displays the numerically evaluated
$\gamma_{k_x,+}^{(\alpha)}$ as a function of $t_2/t_1$ for fixed values of
$B_s=t_1$ and $2t_2$. In the $B_s-t_2$ plane, $\gamma_{k_x,+}^{(\alpha)}$'s
are calculated along the dashed arrows as shown in the left-hand side of
Fig. \ref{fig_MI_Bs}. With increasing $t_2/t_1$ for a fixed $B_s$, we find
that $\gamma_{k_x,+}^{(\alpha)}$ for $k_x=0 (\pi)$ jumps from zero to $\pi$,
like a step function, when $t_1/t_2$ crosses $B_s=B_{s1} (B_{s2})$. Solving
$B_s=B_{s1}$ or $B_{s2}$ with respect to $t_2/t_1$ using Eqs. (\ref{eq_Bs1})
and (\ref{eq_Bs2}), the critical values of $t_1/t_2$ for $k_x=0$ and $\pi$ are
given by 
\begin{eqnarray}
\left(\frac{t_2}{t_1}\right)_{c1}&=&\sqrt{1+\left(\frac{B_s}{2t_1}\right)^2}-\frac{B_s}{2t_1},\\
\left(\frac{t_2}{t_1}\right)_{c2}&=&\left(\frac{B_s}{t_1}\right)^{-1},
\end{eqnarray}
respectively. These critical values are indicated by the vertical arrows in
Fig. \ref{fig_gamma} for $B_s=t_1$ and $2t_1$. Eventually, we can reproduce
the MI phase diagram shown in Fig. \ref{fig_MI_Bs} by looking at
$\gamma^{(\alpha)}_{k_x\sigma}$ for $k_x=0$ and $\pi$. 

When $k_x$ is fixed at an intermediated value between 0 and $\pi$, we have
numerically found that $\gamma^{(\alpha)}_{k_x,+}=0$ for a given parameter set
of $(t_1,t_2,B_s)$, satisfying $k_x<k_c(t_1,t_2,B_s)$ and
$\gamma^{(\alpha)}_{k_x,+}=\pi$ otherwise. At $(t_1,t_2,B_s)=(1,0.8,1)$, for
example, $k_c(t_1,t_2,B_s)\simeq 0.53\pi$, see Eq. (\ref{eq_kc}) and
Fig. \ref{fig_disp_Bs}(d). Therefore, $\gamma^{(\alpha)}_{0.53\pi,+}$ behaves
like $\pi\cdot\theta(t_2/t_1-0.8)$ as a function of $t_2/t_1$ for $B_s=t_1$ or
like $\pi\cdot\theta(B_s/t_1-1)$ as a function of $B_s/t_1$ for $t_2/t_1=0.8$,
where $\theta(x)$ is a step function. In other words, the AFI' (including the
NI at $B_s=0$) and AFI phases are, respectively, characterized by
$\gamma_{k_x\sigma}^{(\alpha)}=0$ and $\pi$ for any value of $k_x$.

The bulk-edge correspondence\cite{halperin,hatsugai2} tells us that the
nontrivial Berry phase $\gamma$ in the AFI suggests an existence of the
zero-energy edge state when making a cut on $t_2$ bonds. Since the wave
function in NI(AFI') and AFI phases are mainly confined within the $t_1$
squares and the $t_2$ bonds, respectively, then different types of edge state
are expected to characterize these two insulating phases. Physics of the
edge state lies in the fact that the antibonding orbitals on $t_2$ bonds are
occupied in the AFI phase and, when making a cut, the dangling bonds
remain. This is a 2D analog of polyacetylene where the cut on a double
covalent bond produces an edge state.\cite{su,ryu} 

\begin{figure}[tbp]
\begin{center}
\includegraphics[width=8cm]{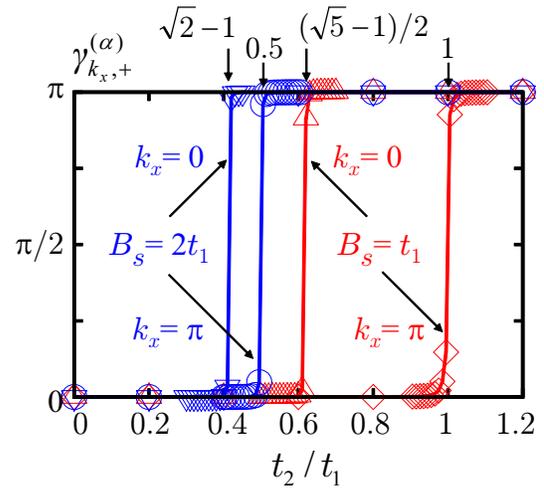}
\caption{(Color online) Berry phase $\gamma_{k_x,+}^{(\alpha)}$ for
  $L_x=L_y=512$ as a function of $t_2/t_1$ for $B_s=t_1$ and $2t_1$ with
  $k_x=0$ and $\pi$. We have checked that $\gamma_{k_x,+}^{(\alpha)}$
  converges to a step function as the system size increases. The positions of
  critical $(t_2/t_1)_c$ are indicated by vertical arrows. 
\label{fig_gamma}}
\end{center}
\end{figure}

\section{summary and concluding remarks}\label{sec_4}
In summary, we have revealed a rich variety of magnetic and M-I phase diagrams
of the 1/5-depleted square-lattice Hubbard model at 1/4 filling within the
mean field approximation. The mean-field phase diagram in $U-t_1$ or $U-t_2$
plane consists of three magnetic phases of nonmagnetic, $\bm Q=\bm 0$ antiferromagnetic, and
completely ferromagnetic type.  The phase transition between nonmagnetic and
antiferromagnetic phases is continuous, which is consistent with the RPA instability
for $t_1<t_2$; the other magnetic transitions to ferromagnetic states are
discontinuous. The SU(3) effective theory and the spectrum analyses under the
external staggered or uniform magnetic fields elucidate the properties of
Lifshitz-type M-I phase transitions in each magnetic phase, which separate the
phase diagram into $3\times2=6$ parts. In particular, the existence of the AF
bond SDW order is shown based on the perturbation theory for small $t_1$,
which corresponds to an extension of the AFI phase for $t_1<t_2$ and $t_1\ll
U$ in the mean-field phase diagram. 
When $U=0$ and $t_1\ll t_2$, the model is
reduced to the 1/2-filled square-lattice tight-binding model of the antibonding
orbitals on the $t_2$ bonds. Then the weak-coupling theory tells us that the perfect 
nesting leads to the AF SDW insulating order at infinitesimal $U$, which corresponds 
to the very small onset of $U/t_2$ toward the AFI phase shown in Fig. \ref{fig_pd}. 
This AFI phase may be regarded as the two-dimensional analog of the $4k_F$ 
charge order discussed in the one-dimensional 1/4-filled Hubbard model 
for the organic conductor (TMTTF)$_2$X.\cite{jerome,ung} 
In this sense, despite the Hartree-Fock treatment of the present study, 
we can say that the emergence of the four distinct phases for the small $U$ region and 
the hidden symmetry in the AFI phase, which is characterized by the nontrivial 
Berry phase, should be robust.

One of the most interesting issues to be addressed is whether the FM and FI
phases appearing in the large $U$ region survive or not after including
effects of electron correlation beyond the Hartree-Fock approximation. As
mentioned for $t_1\ll t_2$ and $U$, the answer is no. In the opposite limit of
$t_2=0,t_1\ll U$, unfortunately, $E_{\rm NI}=-2\sqrt{2}t_1 < E_{\rm FM}=-2t_1$
always holds by solving a 4-site Hubbard model with two electrons for no double
occupancy. Then what about the case of $t_1\simeq t_2\ll U$? To partly answer
this question, we have applied numerical exact diagonalization for the 16-site
cluster shown in Fig. \ref{fig_lattice}(a) with eight electrons. For any of open,
periodic, and antiperiodic boundary conditions, as a result, the
ground-state energy is always found in the spin-singlet sector for $U$ at
least smaller than 20 at $t_1=t_2=1$. We need much more sophisticated
treatments to settle this issue, which is beyond the scope of this paper. 

The Dirac-cone and flat-band dispersions are characteristic of the present
model, and the lifting degeneracy of the SU(3) multiplet rules the physics
around the SU(3) point. The SU(3) effective Hamiltonian can be constructed
solely by the symmetry arguments and, therefore, the present peculiar band
structure may be relevant to some realistic cases. For example, complex
hopping matrices, like the Haldane model\cite{haldane} (due to a magnetic flux) or
Kane-Mele model\cite{kane} (spin-orbit couplings), introduce a $\lambda_2$ term
and a staggered modulation in $t_1$'s, due to a lattice symmetry lowering from
the original $\rm C_4$ to $\rm C_2$, bringing a $\lambda_3$ term in the SU(3)
model. 

The AFM phase is rather specific in that the flat band along $\Gamma$-X does
not hybridize with the crossing band at $k_x=k_c$ because of the different
parity symmetry of these two states with respect to exchanging B and D
sites.\cite{tsunetsugu} This parity protected AFM phase separates the NI and
AFI phases, which are classified according to the different Berry phases. This
is the manifestation of the fact that topologically different insulating
phases can not be connected unless the charge gap closes at the junction. 

\begin{acknowledgments}
The authors would like to thank H. Tsunetsugu, M. Oshikawa, I. Maruyama, and
K. Ohgushi for valuable comments and discussions. This work has been supported
by a Grant-in-Aid for Scientific Research on Innovative Areas ``Heavy
Electrons'' (Grant No. 20102008) and (C) (Grant No. 25400357) 
from the Ministry of Education, Culture, Sports, Science and Technology, Japan.
This work was carried out
by the joint research in the Institute for Solid State Physics, the University
of Tokyo. 
\end{acknowledgments}



\begin{thebibliography}{99}
\bibitem{comment4} Here, AF (F) stands for the staggered (uniform) spin polarization under the $B_s$ ($B_u$) field. We use this terminology for the external field induced phase to be consistent with the mean-field phase diagram shown in Fig. \ref{fig_pd}.
\bibitem{comment1} Since $t_2$ is positive definite, the antibonding orbital is located lower in energy.
\end{thebibliography}
\end{document}